# Dynamic Asymmetric Causality Tests with an Application

Abdulnasser Hatemi-J

College of Business and Economics, UAE University

Email: AHatemi@aueu.ac.ae

(arXiv:2106.07612)

## Abstract

Testing for causation—defined as the preceding impact of the past value(s) of one variable on the current value of another one when all other pertinent information is accounted for—is increasingly utilized in empirical research of the time-series data in different scientific disciplines. A relatively recent extension of this approach has been allowing for potential asymmetric impacts since it is harmonious with the way reality operates in many cases (Hatemi-J, 2012). The current paper maintains that it is also important to account for the potential change in the parameters when asymmetric causation tests are conducted, as there exists a number of reasons for changing the potential causal connection between variables across time. The current paper extends therefore the static asymmetric causality tests by making them dynamic via the usage of subsamples. An application is also provided consistent with measurable definitions of economic or financial bad as well as good news and their potential causal interaction across time.

## 1. Introduction

From cradle to grave, one of the most prevalent and persistent questions in life is figuring out what is the cause and what is the effect when certain pertinent events that take place are observed. This subject must have been one of the most inspirational and important issues since the dawn of humankind. Throughout the history, many philosophers have devoted their pondering to causality as an abstraction. Yet there is no common definition of causality and above all, there is no common or universally accepted approach for detecting or measuring causality. Since the pioneer notion of Wiener (1956) and the seminal contribution of Granger (1969), testing for the predictability impact of a variable on another one has increasingly gained popularity and practical usefulness in different fields when the variables are quantified across time. This approach is known as Granger causality in the literature and it describes a situation in which the past values of one variable (i.e. the cause variable) are statistically significant in an autoregressive regression model of another variable (i.e. the effect variable) when all other relevant information is also accounted for. The null hypothesis is defined as zero restrictions imposed on the parameters of the cause variable in the autoregressive model when the dependent variable is the potential effect variable. If the null hypothesis is not accepted empirically, it is taken as an empirical evidence for causality in the sense of Wiener-Granger.[1] There have been several extensions of this method especially since the discovery of unit roots and stochastic trends, which is a common property of many time-series variables that quantify economic or financial processes across time. Granger (1986, 1988) as well as Engle and Granger (1987) suggested testing for causality via an error correction model if the variables are integrated. Toda and Yamamoto (1995) proposed a modified Wald (1939) test static in order to take into account the impact of unit roots when causality tests are conducted within the vector autoregressive (VAR) model by adding additional unrestricted lags. Hacker and Hatemi-J (2006, 2012) suggested bootstrapping with leverage adjustments in order to generate accurate critical values for the modified Wald test since the asymptotical ones are not precise when the desirable statistical assumptions for a good model are not satisfied according to the conducted Monte Carlo simulations by the authors. The bootstrap corrected tests appear to have better size and power properties compared to the asymptotic ones, especially in the small sample sizes.

However, there are numerous reasons for the potential causal connection between the variables to have an asymmetric structure. It is commonly agreed in the literature that markets with asymmetric information prevail (based on the seminal contributions of Akerlof, 1970;

---

[1] There exists alternative designs such as Sims (1972) and Geweke (1982).

Spence, 1973; and Stiglitz, 1974). People frequently react stronger to a negative change in contrast to a comparable positive one.[2] There are also natural restrictions that can lead to the asymmetric causation phenomenon. For instance, there is a limit on the potential price decrease of any normal asset or commodity since the price cannot drop below zero. However, there is no restriction on the amount of the potential price increase. In fact, if the price decreases by a given percentage point and then increases again by the same percentage point; it will not end up with the initial price level but at a lower level. This is true even if the process occurs in the reverse order. There are also moral and/or legal limitations that can lead to an asymmetric behavior. For example, if a company manages to increase its profit by the $P\%$ at a given period, it is feasible and easy to expand the business by that percentage point. However, if the mentioned company experiences a loss by the $P\%$, it is not that easy to implement an immediate contraction of the business operation by the $P\%$. The contraction is usually less than the $P\%$ and it can take longer time to be realized compared to the corresponding expansion. It is naturally easier for the company to hire people than firing them. In the markets for many commodities, it can also be observed clearly that there is an inertia effect for the price decreases compared to the price increases. Among others, the fuel market can be mentioned. When the oil price increases, there seems to be an almost immediate price increase of the fuel prices and by the same proportion if not more. However, when the oil price decreases there is a lag in the price decrease of the fuel prices and the adjustment might not be implemented fully. This indicates that the fuel prices adjustments are asymmetric with regard to the oil price changes under the *ceteris paribus* condition. In order to account for these kind of potential asymmetric causation in the empirical research based on time-series data, Hatemi-J (2012) suggests implementing asymmetric causality tests that the author introduces. However, these asymmetric causality tests are static by nature.

  The objective of the current paper is to extend these asymmetric causality tests to a dynamic context by allowing for the underlying causal parameters to vary across time, which can be achieved by using subsamples. There are several advantages for using this dynamic parameter approach. One of the advantages of the time-varying parameter approach is that it takes into account the well-known Lucas (1976) critique, which is an essential issue from the policy-making point of view. Peoples' preferences can change across time that result in a change in their behavior and thereby changing the economic or financial process. There are path-breaking technological innovations and progresses that happen with time. Major organizational restructuring can take place across time. Unexpected major events, such as the current COVID-19 pandemic, can occur. All these events can result in a change in the potential causal connection between the underlying variables in a model. Thus, a dynamic parameter model can be more informative and it can better present the way things operate in reality. From a correct model specification perspective also, the dynamic parameter approach can be preferred to the constant parameter approach in addition to being more informative. Since the dynamic testing of the potential causation connection is more informative than the static approach, it can shed light on the extent of the pertinent phenomenon known as the correlation risk in the financial literature. According to Meissner (2014) correlation risk is defined as the potential risk that the strength of the relationship between financial assets varies unfavorably across time. This issue can have crucial ramifications for investors, institutions and the policy makers.

---

[2] According to Longin and Solnik (2001), Ang and Chen (2002), Hong and Zhou (2008) and Alvarez-Ramirez et al. (2009), among others, there is indeed an asymmetric behavior by the investors in the financial markets since they have a tendency to respond stronger to the negative news compared to the positive ones.

The subsequent definitions are utilized in this paper.

*Definition 1.* A *n*-dimensional stochastic process $(x_t)_{t=1,\cdots,T}$ measured across time is integrated of degree *1*, signified as I(*1*), if it must be differenced *once* for becoming a stationary process. That is, $x_t$~I(*1*) if $\Delta x_t$~I(*0*), where the denotation $\Delta$ is the first difference operator.

*Definition 2.* Define $(\varepsilon_t)_{t=1,\cdots,T}$ as a *n*-dimensional stochastic process. Thus, during any time period $t \in \{1, \cdots, T\}$, the positive and negative shocks of this random variable $\varepsilon_t$ (i.e. $\varepsilon_t^+$ and $\varepsilon_t^-$) are identified as the following:

$$\varepsilon_t^+ := max(\varepsilon_t, 0) := (max(\varepsilon_{1t}, 0), \cdots, max(\varepsilon_{nt}, 0)) \qquad (1)$$

and

$$\varepsilon_t^- := min(\varepsilon_t, 0) := (min(\varepsilon_{1t}, 0), \cdots, min(\varepsilon_{nt}, 0)) \qquad (2)$$

The definition of the positive and negative shocks were suggested by Granger and Yoon (2002) for testing for hidden cointegration.[3]

The rest of the paper is organized as follows. Section 2 introduces the methodology of the dynamic asymmetric causality testing. Section 3 provides an application of the potential causal impact of the oil prices on the world's largest stock market accounting for rising and falling prices using both the static and the dynamic asymmetric causality tests. Conclusions are offered in the final section.

---

[3] Hatemi-J (2020a) has extended this method for testing for hidden panel cointegration. For asymmetric panel causality tests see Hatemi-J (2020b).

## 2. Dynamic Asymmetric Causality Testing

The implementation of the causality tests in the sense of Wiener-Granger is operational within the vector autoregressive (VAR) model of Sims (1980). The asymmetric version of this test method is introduced by Hatemi-J (2012).[4] Consider the following two I(*1*) variables with deterministic trend parts:[5]

$$x_{1t} = a + bt + x_{1t-1} + \varepsilon_{1t}, \tag{3}$$

and

$$x_{2t} = c + dt + x_{2t-1} + \varepsilon_{2t}, \tag{4}$$

where *a*, *b*, *c* and *d* are parametric constants and *t* is the deterministic trend term. The positive and negative partial sums of the two variables can be recursively defined as the following based on the definitions of shocks presented in equations (1) and (2):

$$x_{1t}^+ := \frac{at + \left[\frac{t(t+1)}{2}\right]b + x_{10}}{2} + \sum_{i=1}^{t} \varepsilon_{1i}^+ \tag{5}$$

$$x_{1t}^- := \frac{at + \left[\frac{t(t+1)}{2}\right]b + x_{10}}{2} + \sum_{i=1}^{t} \varepsilon_{1i}^- \tag{6}$$

$$x_{2t}^+ := \frac{ct + \left[\frac{t(t+1)}{2}\right]d + x_{20}}{2} + \sum_{i=1}^{t} \varepsilon_{2i}^+ \tag{7}$$

$$x_{2t}^- := \frac{ct + \left[\frac{t(t+1)}{2}\right]d + x_{20}}{2} + \sum_{i=1}^{t} \varepsilon_{2i}^- \tag{8}$$

---

[4] Bahmani-Oskooee et al. (2016) extend the test to the frequency domain.
[5] For the simplicity of expression we assume that *n*=2. However, it is straightforward to generalize the results.

Where $x_{10}$ and $x_{20}$ are the initial values. Note that the required conditions of having $x_{1t} = x_{1t}^+ + x_{1t}^-$ and $x_{2t} = x_{2t}^+ + x_{2t}^-$ are fulfilled.[6] Interestingly, the values expressed in equations (5)-(8) have also economic or financial implications in terms of measuring good or bad news that can affect the markets. It should be mentioned that the issue of whether to have the deterministic trend parts in the data generating process for a given variable is an empirical issue. In some cases, there might be need for both a drift and a trend and in other cases, it might be sufficient with a drift without any trend. It is also possible to have no drift and no trend. For the selection of the deterministic trend components, the procedure suggested by Hacker and Hatemi-J (2010) can be useful.

The asymmetric causality tests can be implemented via the vector autoregressive model of order $p$ as introduced originally by Sims (1980), i.e. the VAR($p$). Let us consider testing for the potential causality between the positive components of these two variables. Then, the vector consisting of the dependent variables is defined as $x_t^+ = (x_{1t}^+, x_{2t}^+)$ and the following VAR($p$) can be estimated based on this vector:

$$x_t^+ = B_0^+ + B_1^+ x_{t-1}^+ + \cdots + B_p^+ x_{t-p}^+ + B_{p+1}^+ x_{t-p-1}^+ + v_t^+ \tag{9}$$

where $B_0^+$ is the $2\times1$ vector of intercepts, $B_r^+$ is a $2\times2$ matrix of parameters to be estimated for lag length $r$ ($r = 1,..., p$) and $v_t^+$ is a $2\times1$ vector of the error terms. An important issue before using the VAR($p$) for drawing inference is to determine the optimal lag order $p$. This can be achieved, among others, by minimizing the information criterion suggested by Hatemi-J (2003), which is expressed as the following:

$$HJC = \ln\left(\left|\widehat{\Pi}_p^+\right|\right) + p\left(\frac{n^2 \ln T + 2n^2 \ln(\ln T)}{2T}\right), \quad p = 1,...,p_{\max}. \tag{10}$$

Where $\left|\widehat{\Pi}_p^+\right|$ is the determinant of the variance–covariance matrix of the error terms in the VAR model that is estimated based on the lag length $p$, ln is natural logarithm, $n$ is the number of time-series included in the VAR model and $T$ is the full sample size used for estimating the parameters in that model.[7] The lag order that results in the minimum value of the information criterion is to

---

[6] For the proof of these results and for the transformation of I(2) and I(3) variables into the cumulative partial sums of negative and positive components see Hatemi-J and El-Khatib (2016).
[7] The Monte Carlo simulations conducted by Hatemi-J (2008) demonstrate clearly that the information criterion expressed in equation (10) is successful in selecting the optimal lag order when the VAR model is used for

be selected as the optimal lag order. It is also important that the off diagonal elements in the variance-covariance matrix are zero. Therefore, tests for multivariate autocorrelation need to be performed in order to verify this issue. The null hypothesis that the *j*th element of $x_t^+$ does not cause the *kth* element of $x_t^+$ can be tested via a Wald (1939) test statistic.[8] The null hypothesis of non-causality can be formulated as the following:

$$H_0: \text{The row } k, \text{column } j \text{ element in } B_r^+ \text{ equals zero for } r = 1,\ldots, p. \quad (11)$$

For a densely representation of the Wald test statistic, we need to make use of the following denotations:[9]

$X^+ := (x_1^+, \ldots, x_T^+)$ as a ($n \times T$) matrix, $D^+ := (B_0^+, B_1^+, \ldots, B_{p+1}^+)$ as a ($n \times (1+n \times (p+1))$) matrix, $Z_t^+ := [1, x_t^+, x_{t-1}^+, \ldots, x_{t-p}^+]'$ as a $((1+n \times (p+1)) \times 1)$ matrix, $Z^+ := (Z_0^+, \ldots, Z_{T-1}^+)$ as a $((1+n \times ((p+1)+1)) \times T)$ matrix and $V^+ = (v_1^+, \ldots, v_T^+$ as a ($n \times T$) matrix. Via these denotations, we can express the VAR model and the Wald test statistic compactly as the following:

$$X^+ = D^+ Z^+ + V^+ \quad (12)$$

$$Wald^+ = (C\beta^+)' \left[ C \left( (Z^{+'}Z^+)^{-1} \otimes \widehat{\Pi}_u^+ \right) C' \right]^{-1} (C\beta^+) \quad (13)$$

The parameter matrix $D^+$ is estimated via the multivariate least squares as the following:

$$\widehat{D}^+ = X^+ Z^{+'} (Z^+ Z^{+'})^{-1} \quad (14)$$

Note that $\beta^+ = vec(\widehat{D}^+)$ and *vec* is the column-stacking operator. That is

$$\beta^+ = vec(\widehat{D}^+) = \left( (Z^+ Z^{+'})^{-1} \otimes I_n \right) vec(X^+) \quad (15)$$

The denotation $\otimes$ is the Kronecker product operator and $C$ is a $((p+1) \times n) \times (1+n \times (p+1))$ indicator matrix that includes 1 elements for restricted parameters and 0 elements for the unrestricted ones under the null hypothesis. $I_n$ is a ($n \times n$) identity matrix. $\widehat{\Pi}_u^+$ represents the variance-covariance matrix of the unrestricted VAR model as expressed by equation (12), which can be estimated as the following:

$$\widehat{\Pi}_u^+ = \frac{\widehat{V_u^+}'\widehat{V_u^+}}{T - q} \qquad (16)$$

---

[8] forecasting purposes. In addition, the simulations show that this information criterion is robust to the ARCH effects and performs well when the variables in the VAR model are integrated. See also Mustafa and Hatemi-J (2021) for more information on this information criterion.

[8] It should be mentioned that the additional unrestricted lag has been added to the VAR model for taking into account the impact of one unit root consistent with the results of Toda and Yamamoto (1995). Multivariate tests for autocorrelation needs also to be implemented to make sure that the off diagonal elements in the variance and covariance matrix are zero. See Hatemi-J (2004) regarding multivariate tests for autocorrelation.

[9] It should be pointed out that this formulation requires that the *p* initial values for each variable in the VAR model are accessible. For the particulars on this requirement, see Lutkepohl (2005).

Note that the constant $q$ represents the number of parameters that are estimated in each equation of the VAR model. By using the presented denotations, the null hypothesis of no causation might also be formulated as the following expression:

$$H_0: C\beta^+ = 0 \qquad (17)$$

The Wald test statistic expressed in (13) that is used for testing the null hypothesis of non-causality as defined in (11) based on the estimated VAR model in equation (12) has the following distribution asymptotically:

$$Wald^+ \xrightarrow{d} \chi_p^2 \qquad (18)$$

This is the case if the assumption of normality is fulfilled. Thus, the Wald test statistic for testing for potential asymmetric causal impacts has a $\chi^2$ distribution with the number of degrees of freedom equal to the number of restrictions under the null hypothesis of non-causality, which is equal to $p$ in this particular case. This result holds for a corresponding VAR model for negative components or any other combinations also. For the proof, see Proposition 1 in Hatemi-J and El-Khatib (2016).

However, if the assumption of the normal distribution of the underlying data set is not fulfilled, the asymptotical critical values are not accurate and bootstrap simulations need to be performed in order to obtain accurate critical values. If the variance is not constant or if the ARCH effects prevail than the bootstrap simulations need to be conducted with leverage adjustments. The size and power properties of the test statistics based on the bootstrap simulation approach with leverage corrections has been investigated by Hacker and Hatemi-J (2006, 2012) via the Monte Carlo simulations. The simulation results provided by the mentioned authors show that the causality test statistic based on the leveraged bootstrapping has correct size and higher power compared to a causality test based on the asymptotic distributions, especially when the sample size is small or when the assumption of normal distribution and constant variance of the error terms are not fulfilled.

The bootstrap simulations can be conducted as the following. First, estimate the restricted model based on regression equation (12). The restricted model imposes the restrictions under the null hypothesis of non-causality. Second, generate the bootstrap data, i.e. $X^{+^*}$, via the estimated parameters from the regression, the original data and the bootstrapped residuals. This means generating $X^{+^*} = \widehat{D}^+ Z^+ + V^{+^*}$. Note that the bootstrapped residuals (i.e. $V^{+^*}$) are created by $T$ random draws with replacement from the modified residuals of the regression. Each of this random draw with replacement has the same likelihood, which is equal a probability of $1/T$. The bootstrapped residuals need to be mean-adjusted in order to ensure that the residuals have zero expected value in each bootstrap sample. This is accomplished via subtracting the mean value of the bootstrap sample from each residual in that sample. Note that the residuals need to be adjusted by leverages in order to make sure that the variance is constant in each bootstrap sample. Next, repeat the bootstrap simulations 10000 times and estimate the Wald test each time.[10] Use these test values in order to generate the bootstrap distribution of the test. The critical value at the $\alpha$ significance level via the bootstrapping (denoted by $c_\alpha^*$) can be acquired by taking the ($\alpha$)th upper quantile of the distribution of the Wald test that is generated via the bootstrapping. The final step is to estimate the Wald test value based on the original data and compare it to the bootstrap critical value at the $\alpha$ level of significance. The null hypothesis of non-causation is rejected at the $\alpha$ significance level if the estimated Wald test value is higher than the $c_\alpha^*$ (i.e. the bootstrap critical value at $\alpha$ level).

In order to account for the possibility of the potential change in the asymmetric causal connection between the variables, these tests can be conducted using subsamples. A crucial issue within this context is to determine the minimum subsample size that is required for testing for the dynamic asymmetric causality. The following formula that is developed by Phillips et al. (2015) can be used for determining the smallest subsample size ($S$):

$$S = \left[ T \left( 0.01 + \frac{1.8}{\sqrt{T}} \right) \right] \tag{19}$$

---

[10] For more information on leverage adjustments in the univariate cases see Davison and Hinkley (1999) and in the multivariate cases see Hacker and Hatemi-J (2006). For asymmetric panel causality tests see Hatemi-J (2020).

Where *T* is the original full sample size. Note that *S* needs to be rounded up.

Two different approaches regarding the subsamples can be implemented for this purpose. The first one is the rolling window approach, which is based on repeated estimation of the model with subsample size of *S* each time and the window is moved forward by one observation each time. That is, we need to estimate the time varying causality for the following subsamples, where each number represents a point in the time:

$$1, 2, 3, \cdots, S$$
$$2, 3, 4, \cdots, (S + 1)$$
$$3, 4, 5, \cdots, (S + 1), (S + 2)$$
$$4, 5, 6, \cdots, (S + 1), (S + 2), (S + 3)$$
$$\vdots$$
$$(T - S + 1), (T - S + 2), (T - S + 3), \cdots, (T - S + S)$$

This means that the first subsample consists of the range covering the first observation to the *S* observation. The next subsample removes the first observation from *S* and adds the one after *S*. This process continues until the full range is covered. For example, assume that *T*=10 and then we have *S*=6 based on equation (19) when *S* is rounded up.[11] Thus, we have the following subsamples (where each number represents the corresponding time):

$$1, 2, 3, \cdots, S = 1, 2, 3, 4, 5, 6$$
$$2, 3, 4, \cdots, (S + 1) = 2, 3, 4, 5, 6, 7$$
$$3, 4, 5, \cdots, (S + 1), (S + 2) = 3, 4, 5, 6, 7, 8$$
$$4, 5, 6, \cdots, (S + 1), (S + 2), (S + 3) = 4, 5, 6, 7, 8, 9$$
$$(T - S + 1), (T - S + 2), (T - S + 3), \cdots, (T - S + S) = (10 - 6 + 1), (10 - 6 + 2), (10 - 6 + 3), (10 - 6 + 4), (10 - 6 + 5), (10 - 6 + 6) = 5, 6, 7, 8, 9, 10.$$

The second method for determining the multiple subsamples is to start with *S* and recursively add an observation to *S* each time for obtaining the next subsample without removing any observation from the beginning. In this approach, the sample size increases by one observation in each subsample until it covers the full range. That is, the size of the first subsample is equal to *S* and the size of the last one is equal to *T*.

---

[11] Obviously, the sample size needs to be bigger normally than 10 observations in the empirical analysis. Here a very small sample size is assumed for the sake of simplicity of the expression.

The next step is to calculate the Wald test statistic for each subsample and produce its bootstrap critical value at a given significance level. Then following ratio can be calculate for each subsample:

$$TVpCV = \frac{Wald\ Test\ Value\ Based\ on\ the\ Given\ Subsample}{Bootstrap\ Critical\ Value\ at\ the\ Given\ Significance\ Level\ and\ Subsample}$$

(20)

Where *TVpCV* signifies the test value per the critical value at a given significance level using a particular subsample. If this ratio is higher than one, it implies that the null hypothesis of no causality is rejected at the given significance level for that subsample. The 5% and the 10% significance levels can be considered. A graphical illustration of (20) for different subsamples can be informative to the investigator in order to detect the potential change of the asymmetric causal connection between the underlying variables in the model.

An alternative method for estimating and testing the time-varying asymmetric causality tests is to make use of the Kalman (1960) filter within a multivariate setting. However, this method might not be operational if the dimension of the model is rather high and/or the lag order is large.

3. **An Application**

An application is provided for detecting the change of the potential causal impact of oil prices and the world's largest financial market. Two indexes are used for this purpose. The first one is the total share prices for all shares for the US. The second index is the global price of Brent crude in US dollars per barrel. The frequency of the data is yearly and it covers the period 1990-2020. The source of the data is the FRED database, which is provided by the Federal Reserve Bank of St. Louis. Let us denote the stock price index by *S* and the oil price index by *O*. The aim of this application is to investigate whether the potential impact of rising or falling oil prices on the performance of the world's largest stock market is time dependent or not. Interestingly, the US market is not only the largest market and its financial market is the most valuable in the word, the U is also the biggest oil producer in the world. These combinations might make the empirical results of this empirical application more useful and general.

The variables are expressed in the natural logarithm format. The unit root test results of the conducted Phillips-Perron test confirm that each variable is integrated of the first order.[12] Firs, the following linear regression is estimated via the OLS technique for the stock price index (i.e. $x_{1t} = lnS_t$):

$$(lnS_t - lnS_{t-1}) = a + bt + \varepsilon_{1t}, \tag{21}$$

Next, the residuals of the above regression are estimated. That is

$$\hat{\varepsilon}_{1t} = (lnS_t - lnS_{t-1}) - \hat{a} - \hat{b}t.$$

Note that the circumflex implies the estimated value. The positive and negative shocks are measured as the following based on the definitions presented in equations (1) and (2):

$$\hat{\varepsilon}_{1t}^+ := max(\hat{\varepsilon}_{1t}, 0) \quad \text{and} \quad \hat{\varepsilon}_{1t}^- := min(\hat{\varepsilon}_{1t}, 0)$$

The positive and negative partial sums for this variable are defined as the following based on the results presented in equations (5) and (6):

$$(lnS_t)^+ := \frac{\hat{a}t + \left[\frac{t(t+1)}{2}\right]\hat{b} + (lnS_0)}{2} + \sum_{i=1}^{t} \hat{\varepsilon}_{1i}^+ \tag{22}$$

$$(lnS_t)^- := \frac{\hat{a}t + \left[\frac{t(t+1)}{2}\right]\hat{b} + (lnS_0)}{2} + \sum_{i=1}^{t} \hat{\varepsilon}_{1i}^- \tag{23}$$

Where $lnS_0$ signifies the initial value of the stock price index in the logarithmic format, which is assumed to be zero in this case. Note that the equivalency condition $lnS_t = (lnS_t)^+ + (lnS_t)^-$ is fulfilled. It should be mentioned that the value expressed by equation (22) represents the good news with regard to the stock market while the value expressed by equation (23) signifies the bad news pertaining to the same market. The oil price index can also be transformed into cumulative partial sums of positive and negative components in an analogous way. Note that a drift and a trend was included in the equation of each variable since it seems to be needed based on the graphs presented in Figure 1.

---

[12] The unit root test results are not reported but they are available on request.

The data set can be transformed by a number of user-friendly statistical software components such as Hatemi-J (2014) in Gauss, Hatemi-J and Mustafa (2016a) in Octave, Hatemi-J and Mustafa (2016b) in Visual Basic Applications (VBA) and El-Khatib and Hatemi-J (2017) in C++.[13] The dynamic asymmetric causality tests based on bootstrap simulations are conducted by statistical software component created by Hatemi-J and Mustafa (2021).

Prior to implementing causality tests, diagnostic tests were implemented. the results of these tests are presented in Table 1, which indicate that the assumption of normality is not fulfilled and the conditional variance is not constant in most cases. Thus, making use of the bootstrap simulations with leverage adjustments is necessary in order to produce reliable critical values. This is particularly the case for subsamples since the degrees of freedom are lower.

Both symmetric and the asymmetric causality tests are implemented in a dynamic setting. The results for the symmetric asymmetric causality tests are presented in Tables 2 and 3 based on the 5% and the 10% significance levels. Based on these results, it can be inferred that the oil price does not cause the stock market price index not even at the 10% significance level. The results are also robust to the choice of the subsamples because the same results are obtained during all subsamples. An implication of this empirical finding is that the market is informationally efficient in the semi-strong form with regard to the oil prices as defined by Fama (1970). However, when the tests for dynamic asymmetric causality are implemented, the results show that an oil price decrease does not cause a decrease in the stock market price index and these results are the same across subsamples even at the 10% significance level (see Tables 5 and 7). Conversely, the null hypothesis that an oil price increase does not cause an increase in the stock market price index is rejected during four subsamples. It is interesting that by using only three fewer observations, the null hypothesis of non-causality would be rejected at the 10% significance in contrast to the result for the entire sample period that does not reject the underlying null hypothesis (see Figure 2 and Table 4).

---

[13] Note that the Gauss software component allows only for the stochastic trend, while the other software components allow for both stochastic and deterministic trends when the underlying variable is transformed into positive and negative parts.

Figure 1. The Time Plot of the Variables along with the Cumulative Partial Components for Positive and Negative Parts.

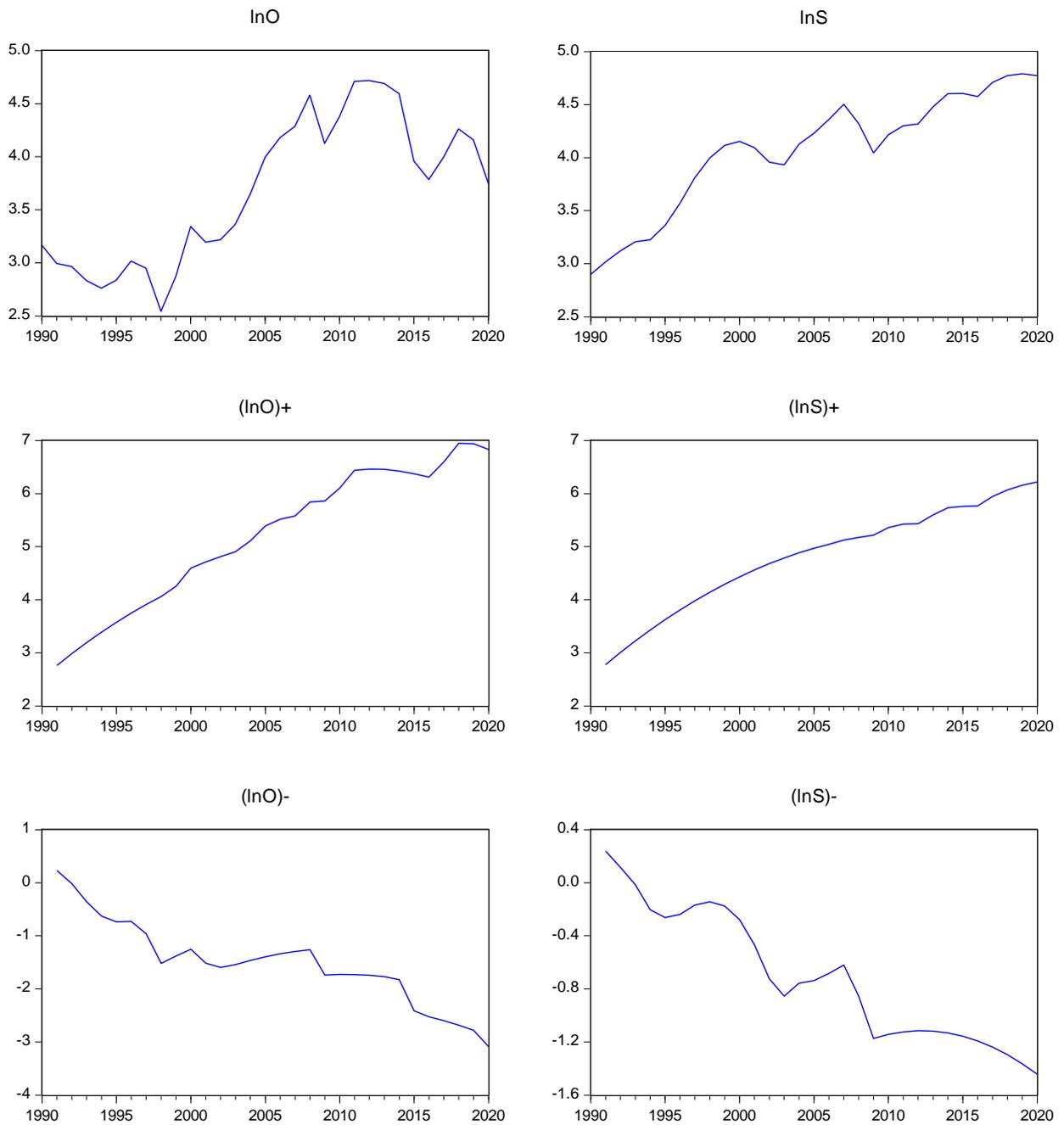

Notes:
The notation ln$O_t$ is representing the oil price index and ln$S_t$ is the US stock market price index for all shares. The corresponding sign indicates the positive and negative components.

Figure 2. The Time Plot of the Causality test Results for the Positive Components at the 10% significance level.

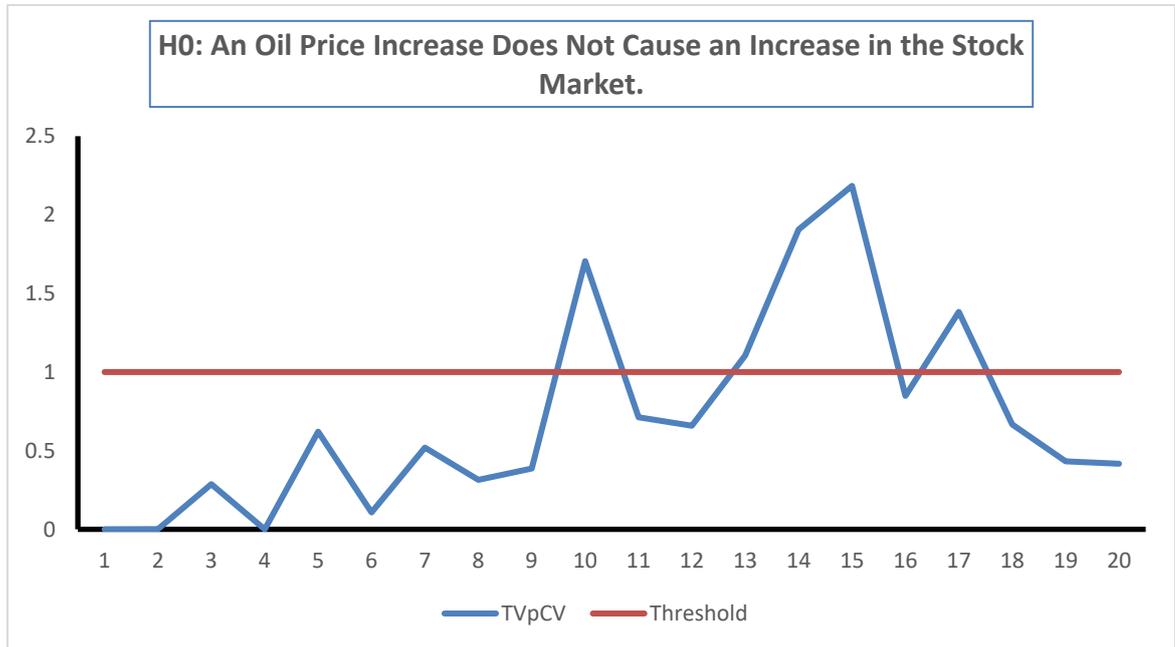

1. Conclusions

Tests for causality in the Wiener-Granger sense are regularly used in empirical research of the time series data in different scientific disciplines. A popular extension of this approach is the asymmetric casualty testing approach as developed by Hatemi-J (2012). However, this approach is static by nature. A pertinent issue within this context is whether the potential asymmetric causal impacts between the underlying variables in a model are steady or not over the selected time span. In order to throw light on this issue, the current paper suggests implementing asymmetric causality tests across time so as to see whether the potential asymmetric causal impact is time depend or not. It is shown how this can be achieved by using subsamples via two different approaches.

An application is provided in order to investigate the potential causal connection of the oil prices with the stock prices of the world's largest financial market within a time-varying setting. The results of the dynamic symmetric causality tests show that the oil prices do not cause the market price index regardless of the subsample size. However, when the dynamic asymmetric causality tests are implemented, the results show that positive oil price changes cause a positive price change in the stock market during certain subsamples. In fact, if only three fewer observations are used compared to the full sample size the results show that there is causality from the oil price increase on the stock market price increase. Conversely, if the full sample size is used (i.e. only three more degrees of freedom), no causality is found. This shows that indeed it can be important to make use of the dynamic causality tests in order to see whether the causality result is robust or not across time.

It should be pointed out that an alternative method for estimating and testing the time-varying asymmetric causality tests is to make use of the Kalman (1960) filter within a multivariate setting. However, this method might not be operational if the dimension of the VAR model is rather high and/or the lag order is large.

The time-varying asymmetric causality tests results can shed light on whether the causal connection between the variables of interest is general or time dependent. This has important practical implications. If the causal connection changes across time then the decision or policy based on this causal impact needs to be time dependent also. This is the case because a static strategy is likely to be inefficient within a dynamic environment.

At the end, the following English proverb that has been quoted by Wiener (1956) says it all.

*"For want of a nail, the shoe was lost;*
*For want of a shoe, the horse was lost;*
*For want of a horse, the rider was lost;*
*For want of a rider, the battle was lost;*
*For want of a battle, the kingdom was lost!"*

**Appendix of Tables**

Table 1. Test Results for Multivariate Normality and Multivariate ARCH in the VAR Model.

| Variables in the Model | The P-value of the Multivariate Normality Test | The P-value of the Multivariate ARCH test |
|---|---|---|
| $[lnS_t, lnO_t]$ | 0.0814 | 0.3743 |
| $[(lnS_t)^+, (lnO_t)^+]$ | 0.5555 | 0.0028 |
| $[(lnS_t)^-, (lnO_t)^-]$ | 0.0087 | 0.0150 |

Notes:
1. *lnO$_t$* signifies the oil price index and *lnS$_t$* is the US stock price index for all shares.. The vector $[(lnS_t)^+, (lnO_t)^+]$ denotes the cumulative partial sum of the positive changes and the vector $[(lnS_t)^-, (lnO_t)^-]$ indicates the cumulative partial sum of the negative changes.
2. The multivariate test of Doornik and Hansen (2008) was implemented for testing the null hypothesis of multivariate normality in the residuals in each VAR model.
3. The multivariate test of Hacker and Hatemi-J (2005) was conducted for testing the null hypothesis of no multivariate ARCH (1).

Table 2: Dynamic Symmetric Causality Test Results at the 5% Significance Level.
($H_0$: The Oil Price Does Not Cause the Stock Market Price Index.)

| SSP | Test Value | 5% Bootstrap CV | TVpCV |
|---|---|---|---|
| 1 | 1.788 | 572.528 | 0 |
| 2 | 1.431 | 52.787 | 0.034 |
| 3 | 0.832 | 20.977 | 0.068 |
| 4 | 0.863 | 13.312 | 0.062 |
| 5 | 0.112 | 10.484 | 0.082 |
| 6 | 0.317 | 4.592 | 0.024 |
| 7 | 0.365 | 5.385 | 0.059 |
| 8 | 0.429 | 5.135 | 0.071 |
| 9 | 0.239 | 4.678 | 0.092 |
| 10 | 0.525 | 4.502 | 0.053 |
| 11 | 0.5 | 4.76 | 0.11 |
| 12 | 0.538 | 4.054 | 0.123 |
| 13 | 0.642 | 4.561 | 0.118 |
| 14 | 0.757 | 4.606 | 0.139 |
| 15 | 0.591 | 4.859 | 0.156 |
| 16 | 0.281 | 4.211 | 0.14 |
| 17 | 0.623 | 3.68 | 0.076 |
| 18 | 0.635 | 4.571 | 0.136 |
| 19 | 0.638 | 4.252 | 0.149 |
| 20 | 0.627 | 4.403 | 0.145 |
| 21 | 0.627 | 4.129 | 0.152 |

DENOTATIONS
------------------------
SSP: The Subsample Period.
CV: The Critical Value.
TVpCV: The Test Value per the Critical Value
TVpCV = (Test Value) / (Bootstrap Critical Value at the Given Significance Level)
If the value of TVpCV > 1, it implies that the null hypothesis of no causality is rejected at the given significance level.

Table 3: Dynamic Symmetric Causality Test Results at the 10% Significance Level.
($H_0$: The Oil Price Does Not Cause the Stock Market).

| SSP | Test Value | 10% Bootstrap CV | TVpCV |
|---|---|---|---|
| 1 | 0.187 | 139.42 | 0.001 |
| 2 | 1.788 | 22.105 | 0.081 |
| 3 | 1.431 | 10.186 | 0.14 |
| 4 | 0.832 | 8.851 | 0.094 |
| 5 | 0.863 | 7.811 | 0.11 |
| 6 | 0.112 | 2.957 | 0.038 |
| 7 | 0.317 | 3.574 | 0.089 |
| 8 | 0.365 | 3.231 | 0.113 |
| 9 | 0.429 | 3.188 | 0.135 |
| 10 | 0.239 | 3.043 | 0.079 |
| 11 | 0.525 | 3.011 | 0.174 |
| 12 | 0.5 | 2.876 | 0.174 |
| 13 | 0.538 | 3.083 | 0.174 |
| 14 | 0.642 | 2.95 | 0.218 |
| 15 | 0.757 | 3.218 | 0.235 |
| 16 | 0.591 | 2.835 | 0.209 |
| 17 | 0.281 | 2.623 | 0.107 |
| 18 | 0.623 | 3.04 | 0.205 |
| 19 | 0.635 | 2.894 | 0.219 |
| 20 | 0.638 | 2.981 | 0.214 |
| 21 | 0.627 | 3.05 | 0.206 |

DENOTATIONS
------------------------
SSP: The Subsample Period.
CV: The Critical Value.
TVpCV: The Test Value per the Critical Value
TVpCV = (Test Value) / (Bootstrap Critical Value at the Given Significance Level)
If the value of TVpCV > 1, it implies that the null hypothesis of no causality is rejected at the given significance level.

Table 4: Dynamic Asymmetric Causality Test Results att the 5% Significance Level.
(H₀: An Oil Price Increase Does Not Cause an Increase in the Stock Market.)

| SSP | Test Value | 5% Bootstrap CV | TVpCV |
|-----|------------|-----------------|-------|
| 1   | 0          | 0.157           | 0     |
| 2   | 0          | 0               | 0     |
| 3   | 0.003      | 0.039           | 0.072 |
| 4   | 0          | 16.548          | 0     |
| 5   | 5.292      | 12.61           | 0.42  |
| 6   | 0.838      | 11.836          | 0.071 |
| 7   | 2.89       | 8.299           | 0.348 |
| 8   | 1.052      | 5.009           | 0.21  |
| 9   | 1.124      | 4.334           | 0.259 |
| 10  | 10.014     | 8.446           | 1.186 |
| 11  | 2.309      | 4.543           | 0.508 |
| 12  | 3.688      | 7.523           | 0.49  |
| 13  | 6.353      | 7.967           | 0.797 |
| 14  | 9.888      | 7.387           | 1.339 |
| 15  | 11.778     | 7.529           | 1.564 |
| 16  | 2.607      | 5.282           | 0.494 |
| 17  | 7.435      | 7.514           | 0.989 |
| 18  | 2.136      | 4.334           | 0.493 |
| 19  | 1.279      | 4.311           | 0.297 |
| 20  | 1.288      | 4.536           | 0.284 |

DENOTATIONS
------------------------
SSP: The Subsample Period.
CV: The Critical Value.
TVpCV: The Test Value per the Critical Value
TVpCV = (Test Value) / (Bootstrap Critical Value at the Given Significance Level)
If the value of TVpCV > 1, it implies that the null hypothesis of no causality is rejected at the given significance level.

Table 5: Dynamic Asymmetric Causality Test Results at the 10% Significance Level.
(H$_0$: An Oil Price Increase Does Not Cause an Increase in the Stock Market.)

| SSP | Test Value | 10% Bootstrap CV | TVpCV |
|-----|------------|------------------|-------|
| 1   | 0          | 0.104            | 0     |
| 2   | 0          | 0                | 0.001 |
| 3   | 0.003      | 0.01             | 0.287 |
| 4   | 0          | 10.351           | 0     |
| 5   | 5.292      | 8.517            | 0.621 |
| 6   | 0.838      | 7.86             | 0.107 |
| 7   | 2.89       | 5.562            | 0.52  |
| 8   | 1.052      | 3.335            | 0.315 |
| 9   | 1.124      | 2.905            | 0.387 |
| 10  | 10.014     | 5.873            | 1.705 |
| 11  | 2.309      | 3.245            | 0.712 |
| 12  | 3.688      | 5.594            | 0.659 |
| 13  | 6.353      | 5.749            | 1.105 |
| 14  | 9.888      | 5.188            | 1.906 |
| 15  | 11.778     | 5.392            | 2.184 |
| 16  | 2.607      | 3.074            | 0.848 |
| 17  | 7.435      | 5.378            | 1.382 |
| 18  | 2.136      | 3.2              | 0.667 |
| 19  | 1.279      | 2.965            | 0.432 |
| 20  | 1.288      | 3.085            | 0.417 |

DENOTATIONS
------------------------
SSP: The Subsample Period.
CV: The Critical Value.
TVpCV: The Test Value per the Critical Value
TVpCV = (Test Value) / (Bootstrap Critical Value at the Given Significance Level)
If the value of TVpCV > 1, it implies that the null hypothesis of no causality is rejected at the given significance level.

Table 6: Dynamic Asymmetric Causality Test Results at the 5% Significance Level.
(H$_0$: An Oil Price Decrease Does Not Cause a Decrease in the Stock Market.)

| SSP | Test Value | 10% Bootstrap CV | TVpCV |
|---|---|---|---|
| 1 | 40.802 | 415.432 | 0.098 |
| 2 | 1.021 | 34.228 | 0.03 |
| 3 | 1.001 | 19.468 | 0.051 |
| 4 | 1.245 | 14.706 | 0.085 |
| 5 | 0.44 | 11.328 | 0.039 |
| 6 | 0.525 | 9.4 | 0.056 |
| 7 | 0.616 | 10.372 | 0.059 |
| 8 | 0.036 | 5.526 | 0.007 |
| 9 | 0 | 5.221 | 0 |
| 10 | 0.353 | 4.948 | 0.071 |
| 11 | 0.362 | 4.86 | 0.075 |
| 12 | 0.373 | 4.692 | 0.08 |
| 13 | 0.382 | 4.387 | 0.087 |
| 14 | 0.387 | 4.717 | 0.082 |
| 15 | 0.372 | 4.664 | 0.08 |
| 16 | 0.081 | 4.374 | 0.019 |
| 17 | 0.085 | 4.338 | 0.02 |
| 18 | 0.087 | 5.029 | 0.017 |
| 19 | 0.084 | 4.757 | 0.018 |
| 20 | 0.078 | 4.499 | 0.017 |

DENOTATIONS
------------------------
SSP: The Subsample Period.
CV: The Critical Value.
TVpCV: The Test Value per the Critical Value
TVpCV = (Test Value) / (Bootstrap Critical Value at the Given Significance Level)
If the value of TVpCV > 1, it implies that the null hypothesis of no causality is rejected at the given significance level.

Table 7: Dynamic Asymmetric Causality Test Results at the 10% Significance Level.
(H$_0$: An Oil Price Decrease Does Not Cause a Decrease in the Stock Market.)

| SSP | Test Value | 10% Bootstrap CV | TVpCV |
|---|---|---|---|
| 1 | 40.802 | 93.964 | 0.434 |
| 2 | 1.021 | 19.269 | 0.053 |
| 3 | 1.001 | 10.783 | 0.093 |
| 4 | 1.245 | 9.204 | 0.135 |
| 5 | 0.44 | 7.935 | 0.055 |
| 6 | 0.525 | 6.316 | 0.083 |
| 7 | 0.616 | 6.836 | 0.09 |
| 8 | 0.036 | 3.601 | 0.01 |
| 9 | 0 | 3.339 | 0 |
| 10 | 0.353 | 3.275 | 0.108 |
| 11 | 0.362 | 3.048 | 0.119 |
| 12 | 0.373 | 3.248 | 0.115 |
| 13 | 0.382 | 3.076 | 0.124 |
| 14 | 0.387 | 3.199 | 0.121 |
| 15 | 0.372 | 2.938 | 0.127 |
| 16 | 0.081 | 2.862 | 0.028 |
| 17 | 0.085 | 2.784 | 0.031 |
| 18 | 0.087 | 3.205 | 0.027 |
| 19 | 0.084 | 2.884 | 0.029 |
| 20 | 0.078 | 3.066 | 0.025 |

DENOTATIONS
------------------------
SSP: The Subsample Period.
CV: The Critical Value.
TVpCV: The Test Value per the Critical Value
TVpCV = (Test Value) / (Bootstrap Critical Value at the Given Significance Level)
If the value of TVpCV > 1, it implies that the null hypothesis of no causality is rejected at the given significance level.